# On the Augmentation of Cognitive Accuracy and Cognitive Precision in Human/Cog Ensembles

Ron Fulbright
Informatics & Engineering Systems
University of South Carolina Upstate
Spartanburg, SC 29303
fulbrigh@uscupstate.edu

**Abstract:** Whenever humans use tools human performance is enhanced. Cognitive systems are a new kind of tool continually increasing in cognitive capability and are now performing high-level cognitive tasks previously thought to be explicitly human. Usage of such tools, known as cogs, are expected to result in ever increasing levels of human cognitive augmentation. In a human/cog ensemble, a cooperative, peer-to-peer, and collaborative dialog between a human and a cognitive system, a human's cognitive capability is augmented as a result of the interaction. The human/cog ensemble is therefore able to achieve more than just the human or the cog working alone. This article presents results from two studies designed to measure the effect information supplied by a cog has on cognitive accuracy, the ability to produce the correct result, and cognitive precision, the propensity to produce only the correct result. Both cognitive accuracy and cognitive precision are shown to be increased by cog-supplied information of different types (policies/rules, examples, and suggestions) and with different kinds of problems (inventive problem solving, and puzzles). Similar effects shown in other studies are compared.

## 1. Introduction

The idea of enhancing human ability with artificial systems is certainly not new. Indeed, human inventions like language, speech, and mathematics have enabled human intellectual, cultural, and social evolution. Likewise, humans create physical tools like wheels, shovels, hammers, and saws to augment human physical ability. Humans also create technology to process information. For example, the abacus was invented thousands of years ago as a calculating device. The abacus, and other calculating devices such as Blaise Pascal's Pascaline in the 1640s [1], allowed people to perform calculations much faster than they could do in their head or by hand on paper. Today, we might use calculators, spreadsheets, and apps. Historically, one of the most important pieces of information technology was the printing press [2]. The printing press gave the average person the ability to copy and distribute information to the masses. The enhanced ability to communicate knowledge transformed religion, science, politics, education, and all human culture.

Along with the industrial revolution came the idea of machines performing or enhancing human cognition. In the 1840s, Ada Lovelace was among the first to envision a machine performing a human cognitive task—musical composition [3]. However, ideas like this were a century before their time. In the 1940s, Vannevar Bush envisioned a system based on microfiche, called the Memex, and discussed how it, through associative linking of information, could enhance a human's ability to store and retrieve information [4].

The most important information technology since the printing press is the modern electronic computer. In the computer age, attention has turned toward making computers able to think like humans (artificial intelligence) [5]. In 1950, Turing discussed whether or not machines themselves could think and offered the "imitation game" as a way to decide if a machine is exhibiting intelligent behavior [6]. In the 1950s, Ross Ashby coined the term *intelligence amplification* maintaining human intelligence could be synthetically enhanced by increasing the human's ability to make appropriate selections on a persistent basis [7].

Inspired by Bush's and Ashby's ideas of human cognitive augmentation, in the 1960s, Licklider and Engelbart were among the first to conceive of how computers could augment human performance. Licklider and Engelbart considered the human and computer as symbiotic components working together as an integrated system [8]. Through the work of Engelbart's Augmentation Research Center, and other groups in the 1950s and 1960s, many of the devices we take for granted today were invented as "augmentation" tools including: the mouse, interactive graphical displays, keyboards, trackballs, WYSIWYG software, email, word processing, and the Internet [9].

Until very recently though, computers and computer software have been nothing more than simple tools assisting a human who still does most of the thinking. However, the last 15 years, has seen development of a new kind of computer system—the *cognitive system*. A cognitive system, or *cog* for short, is a system able to mimic, perform, or replace parts of higher-level human thinking. [10-13] In the spirit of Engelbart, cogs are not intended to replace humans but are meant to augment human cognition [14].

Recently, we have studied the implications of one or more humans interacting in a peer-to-peer collaboration with one or more cogs in what we call a *human/cog ensemble* as shown in Figure 1 [15].

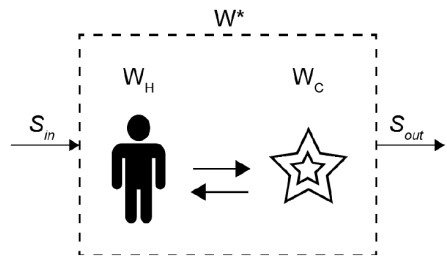

**Figure 1**. A Human/Cog Ensemble

Following Licklider and Engelbart, the human and cog in a human/cog ensemble are engaged in a cooperative, collaborative, and even symbiotic relationship working as a team. The human/cog ensemble accepts information as an input ($S_{in}$), performs cognitive processing and produces information as an output ($S_{out}$). Working together on a task, the human component of a human/cog ensemble performs some of the cognition required ($W_H$) what Engelbart called explicit-human-processes and the cog performs some of the cognition ($W_C$) what Engelbart called explicit-artifact processes. Information travels from human to cog and from cog to human as they collaborate on the task (Engelbart's composite processes). In performing the task, the human/cog ensemble achieves a total amount of cognitive work ($W^*$). Of course, the whole point to this is a human and cog working together can outperform a human working alone ($W^* > W_H$). When this happens, we say the human is cognitively augmented.

A challenge has been to decide how to measure cognitive augmentation. Toward this end we have earlier proposed two metrics, *cognitive accuracy* and *cognitive precision* [16]. Accuracy and precision are well-known concepts in science and engineering. ISO 5725 defines "accuracy" by how close a result is to the true value and "precision" is how close a group of results are to each other [17, 18]. As shown in Figure 2, we define cognitive accuracy of a human/cog ensemble as being able to yield the correct, or desired, result and cognitive precision as being able to yield only the correct, or desired, result [19-21].

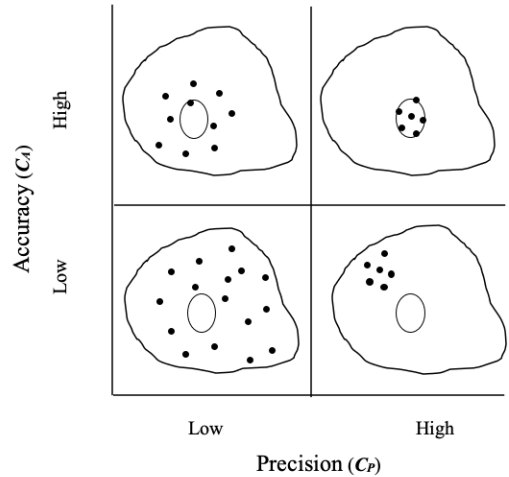

**Figure 2.** Cognitive Accuracy and Cognitive Precision
(The oval represents the correct result)

When working on a task, a human/cog ensemble seeks to produce the correct result every time achieving both high cognitive accuracy and cognitive precision. However, this is not always the case. Sometimes a human/cog ensemble finds a "good" solution (not necessarily the best) and other times, it takes extended time and effort to arrive at a correct result. Variables in cognitive performance include the human's ability to solve the problem at hand, the cog's ability to solve the problem at hand, the nature and quality of the interaction between human and cog, and the human's ability to utilize information supplied by the cog. By comparing how accurate and precise a human working alone is to how accurate and precise a human is while working with a cog, we can measure human cognitive augmentation. Thus, we state our hypothesis as:

> **Hypothesis 1.** A human's cognitive ability while working on a task in collaboration with a cognitive system (a human/cog ensemble) is augmented by the information the cognitive system contributes to the dialog. We expect to see increases in both cognitive accuracy and cognitive precision because of, and as evidence of, the cognitive augmentation.

## 2. Materials and Methods

This paper reports on two experiments in cognitive accuracy and cognitive precision we conducted and compares results with real-world studies done by others. In each of our studies, we presented groups of human problem solvers with a challenge. To simulate the effect of working with a cog, we supplied some participants information they might have received from a cog had they been working with one. To remove any biases and other human/computer interaction factors, we did not actually create and use a cognitive system for these studies. Thereby, our studies focus only on the effect of the *information* a cog would supply in a human/cog collaboration so as not be polluted with effects from how the cog is implemented, how the cog interacts with the human, or how adept the human is at using the cog—all interesting research questions to explore in other studies.

### 2.1. Study #1 – Innovation Challenge

In Study #1, an innovation problem was given to students registered for the INFO 307: Systematic Innovation course at the University of South Carolina Upstate. INFO 307 teaches students an innovation methodology called I-TRIZ. To not bias results with the effect of learning the I-TRIZ methodology, this test was done on the first day of classes over two consecutive semesters before any instruction in the methodology took place. One semester consisted of 25 students and the second semester consisted of 21 students for a total of N=46 participants. Students were given ten minutes to write down as many solutions as they could think of to the following problem:

> *Skeet shooting is a recreational and competitive activity where participants, using shotguns, attempt to break clay targets mechanically launched into the air from fixed stations at high speed and a variety of angles. The problem is the shattered skeet litter the grass field harming the grass. As you know, grass needs sun, water, and nutrients to be healthy. The skeet fragments prevent water and sun from reaching the grass and diminish the healthy nutrients in the soil after they eventually dissolve.*

The I-TRIZ methodology utilizes a database of several hundred innovative concepts, called *operators* gleaned from the study of millions of patents. Each operator serves as a suggestion on how to solve the problem at hand. Practitioners typically use several operators as guidelines to synthesize innovative solutions. One-third of the students were given the above problem statement without any other information at all (no operators). One-third of the students were given the problem statement and a list of five operators (OPS 1), and one-third of the students were given the problem statement and a list of five additional operators (OPS 2) for a total of ten operators. The operators given were:

**OPS 1 =**
1. exclude the source
2. use a disposable object
3. apply liquid support/Introduce a liquid
4. inversion (apply the opposite)
5. phase transformation (freeze/melt/boil; solid/liquid/gas/plasma)

**OPS 2 =**
1. transform a substance to a fluid state
2. self-healing (system corrects itself)
3. formation of mixtures
4. transform the aggregate state to eliminate a harmful effect
5. resources - modified water

These operators were chosen for a reason. The skeet shooting problem has been used as a training/educational aid for many years and as a result has been solved many hundreds of times with a variety of different solutions. Solutions, generally fall into three distinct categories:

**F:** modifying the field/cleaning up the field
**T:** modifying the target (skeet)
**G:** modifying the gun or bullet.

Solutions in the field category (**F**) include: covering the field with a tarp or net, various ways of cleaning up the field, and relocating to a location without a grass field. Solutions in the target category (**T**) include: using targets made of biodegradable material, using targets made of fertilizer and other nutrients good for the grass, and enhancing the rapid dissolvability of the targets. Solutions in the gun category (**G**) include: using a different kind of gun or bullet, and using a laser or electromagnetic gun/target.
For this study, the **T** (modifying the target) category of solutions was chosen as the preferred type of solution because solutions in the other two categories are considered obvious solutions (the first ones most people think of off the top of their head). Operators in OPS1 and OPS2 shown above were chosen with the intent of driving student thinking toward solutions in the **T** category. Cognitive accuracy in this case is measured by the propensity to arrive at a category **T** solution and cognitive precision is measured by the propensity to arrive at only category **T** solutions. Cognitive augmentation is demonstrated by an increase in cognitive accuracy or cognitive precision, or both.

## 2.2. Study #2 – Puzzle Solution

In Study #2, participants were asked to solve four different puzzles listed below and shown in Figure 3.

- Puzzle 1: "Square"  (3-row math puzzle)
- Puzzle 2: "X puzzle"  (diagonal math puzzle)
- Puzzle 3: "4 X 4"  (4-row sequence puzzle)
- Puzzle 4: "Message"  (6-word decryption puzzle)

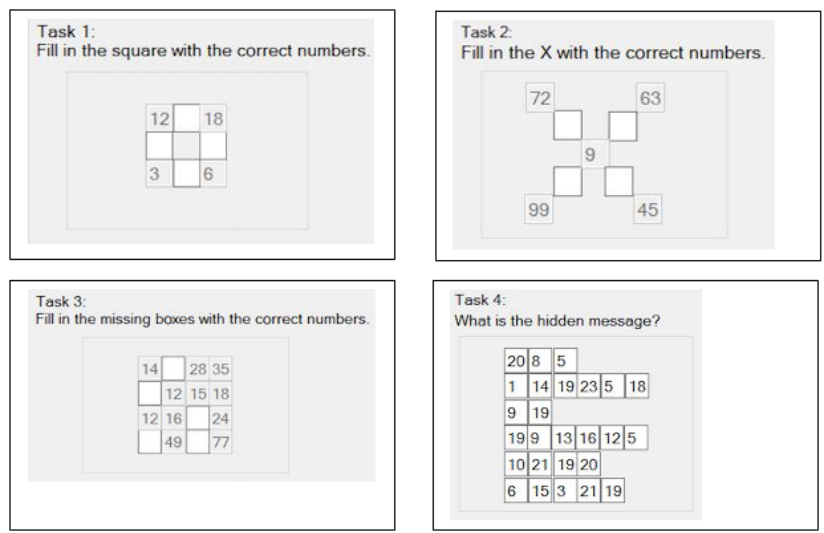

**Figure 3.** Four puzzles participants were asked to solve

The puzzles were presented to the participants one at a time with the participant allowed to continue to the next puzzle only upon successful completion of the current puzzle. Two of the four puzzles involved basic mathematical functions (addition, subtraction, multiplication). One puzzle involved recognizing a pattern in a sequence of numbers. One puzzle involved solving decoding a simple substitution cyber. Each puzzle involved non-trivial kinds of cognition but was simple enough to be solved by anyone with grade-school education and knowledge.

To investigate the effect of different types of information on cognitive performance, some participants were presented with a hint along with the puzzle. One-third of the participants were given no hint (the "normal" group) and served as the control group. One-third of the participants were given a hint in the form of example of a completed puzzle (the "concept" group). The remaining one-third of the participants were given a hint in the form of a guideline or rule as shown below (the "policy" group).

- Square   "Each row is a different mathematical operation."
- X puzzle  "The middle box and the empty box combine to equal the third box."
- 4 X 4    "Each row is based on a specific number. One row is a combination of the other three rows."
- Message  "Each number is tied to a specific letter in the English alphabet."

Participants were given up to one hour to complete the puzzles. If, after an hour, all puzzles were not solved the attempt was counted as a failure. Participants were allowed to submit an attempted solution to a puzzle and then receive a message whether the solution was correct. If incorrect, the participant was allowed to repeat and submit another solution. Attempted solutions were limited to 25. If after 25 attempts the puzzles were not solved, the attempt was listed as a failure. Performance of the participants was assessed in several ways:

- Failure Percentage (inability to solve a puzzle)
- Total Overall Time (total time taken working on the puzzles)
- Average Attempts Per Puzzle

Cognitive accuracy in this study is measured by the success/failure percentage and cognitive precision is measured by the average attempts per puzzle because a lower number of attempts indicates the participant is producing a fewer number of incorrect, or undesired, results. Cognitive augmentation will be demonstrated by an increase in cognitive accuracy or cognitive precision, or both.

## 3. Results

### 3.1. Study #1 – Innovation Challenge

For Study #1, each student solution (N=46) was categorized into one of the three solution categories **F**, **G** or **T** described earlier. The goal was to influence student solutions toward the desired category **T** by supplying guidance in the form of specific I-TRIZ operators. Students receiving a list of operators were more likely to arrive at the desired type of solution than students not receiving any additional information at all as shown in Figure 4. The OPS 1 set of operators increased the number of solutions in the target category by 48% (from 27% to 40%). The OPS 1 and OPS 2 set of operators increased the number of solutions in the target category by 74% (from 27% to 47%). Since producing a solution in the target category is a direct measure of accuracy, the increases noted above are equivalent to cognitive accuracy augmentation as shown in Table 1.

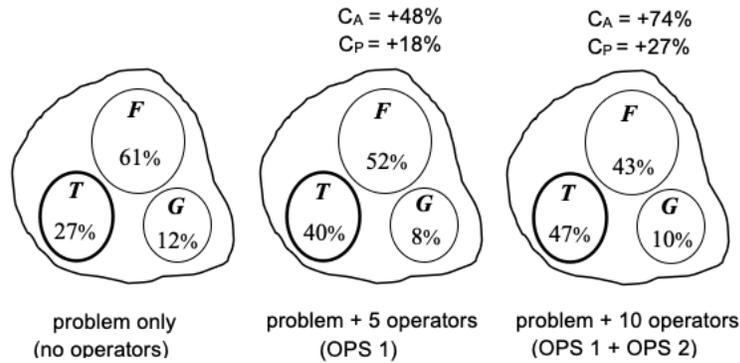

**Figure 4.** Cognitive Accuracy and Cognitive Precision Augmentation Resulting from use of I-TRIZ Operators in Inventive Problem Solving

Since the number of solutions in non-target categories was reduced, cognitive precision was increased by 18% and 27% respectively for OPS 1 and OPS 1 + OPS 2 as shown in Table 1.

**Table 1.** Increase in cognitive accuracy and cognitive precision using I-TRIZ operators

| Group | Δ Cognitive Accuracy | Δ Cognitive Precision |
|---|---|---|
| No Operators | | |
| OPS 1 | $\Delta\, C_A = 48\%$ | $\Delta\, C_P = 18\%$ |
| OPS 1 + OPS 2 | $\Delta\, C_A = 74\%$ | $\Delta\, C_P = 27\%$ |

Unaided, nearly three out of four solutions were in category **F** and category **G** (the non-target, obvious solutions) and only one out of four were in category **T** (the target solution category). However, being cognitively augmented by the OPS 1 and OPS 2 sets of operators resulted in nearly one-half of the solutions being in the target category. This shows the operators had the intended effect of influencing student thinking in the desired direction.

### 3.2. Study #2 – Puzzle Solution
In Study #2, we sought to enhance the puzzle-solving capability of participants by supplying them with two different types of information. Participants were two times more likely to solve the puzzles after receiving "policy" information (rules for the puzzle) and three times more likely to solve the puzzles after receiving "concept" information (an example of a completed puzzle) as shown in Figure 5.

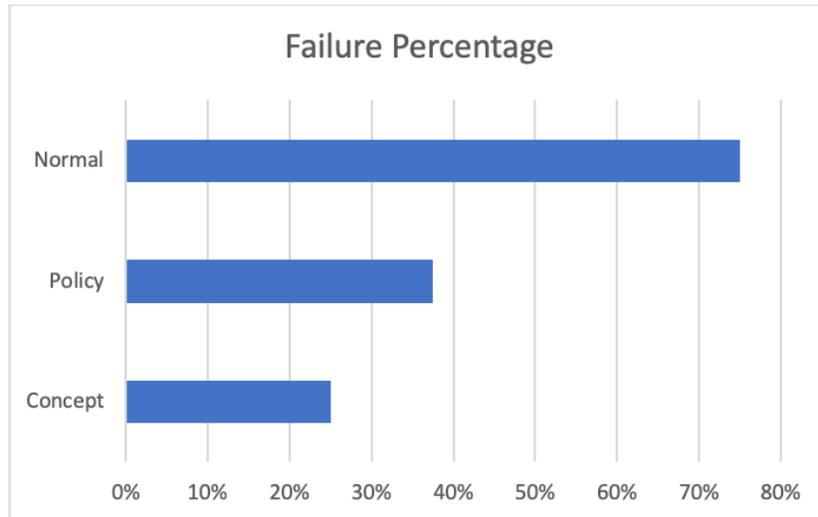

**Figure 5.** Cognitive Accuracy and Cognitive Precision Augmentation

In this study, correctly solving the puzzles is the correct, or desired, result. Unaided participants failed about 75% of the time. Participants augmented with policy information failed only about 37% of the time. Participants augmented with conceptual information failed only about 25% of the time. Therefore, both types of information resulted in increased cognitive accuracy (Δ $C_A$(policy) = 100% and Δ $C_A$(conceptual) =200%).

The same pattern was observed when looking at the total time taken by the participants as shown in Figure 6. Unaided participants took twice as long as participants augmented with policy information and took three times as long as participants augmented with conceptual information.

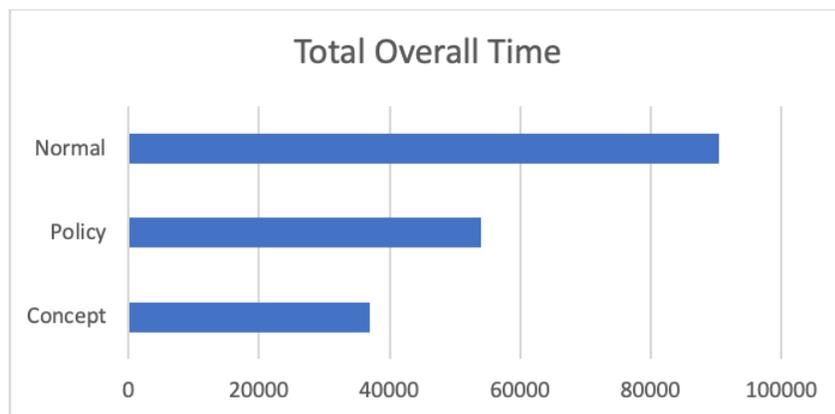

**Figure 6.** Total Overall Time (in seconds)

Since participants were able to attempt each puzzle multiple times, the number of attempts for each puzzle is a measure of precision. Solving the puzzle on the first attempt (as some participants did) would be as precise as possible. Having to attempt to solve the puzzle multiple times before getting it correct is analogous to missing the bullseye multiple times before finally hitting it when throwing darts.

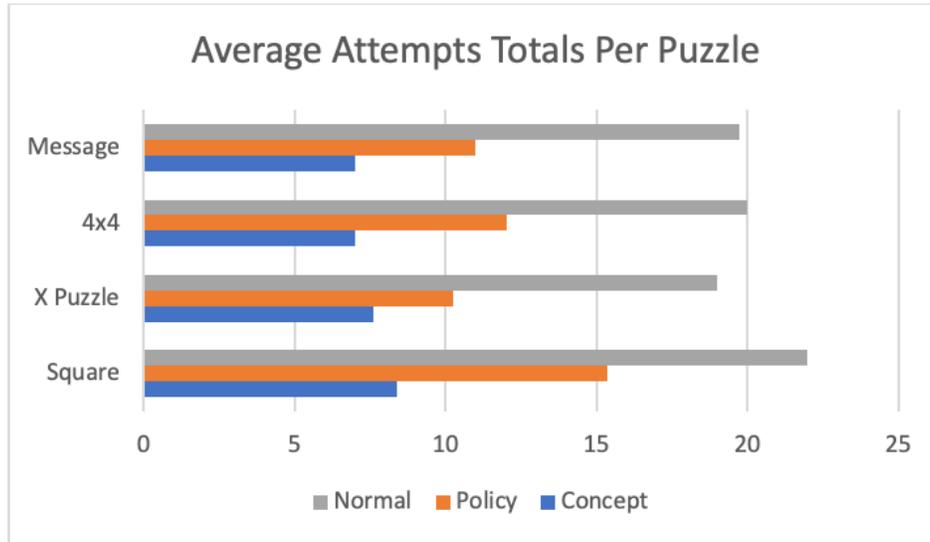

**Figure 7.** Average Attempts per Puzzle

As shown in Figure 7, unaided participants averaged 18-22 attempts at the puzzles before finally correctly solving them. Participants augmented with policy information averaged 11-16 attempts. Participants augmented with conceptual information averaged only 7-8 attempts. These results yield increases in cognitive precision as shown in Table 2.

**Table 2.** Increase in cognitive precision based on information type

| Puzzle | Policy Information | Conceptual Information |
|---|---|---|
| Message | $\Delta\, C_P = 42\%$ | $\Delta\, C_P = 63\%$ |
| 4x4 | $\Delta\, C_P = 40\%$ | $\Delta\, C_P = 65\%$ |
| X | $\Delta\, C_P = 40\%$ | $\Delta\, C_P = 65\%$ |
| Square | $\Delta\, C_P = 27\%$ | $\Delta\, C_P = 64\%$ |

Cognitive precision was increased by 27%-42% using policy information (rule for the puzzle). The "square" puzzle is an outlier because of the nature of that puzzle being a simple cypher substitution. Cognitive precision was increased by 63%-65% using conceptual information (examples of completed puzzles).

## 4. Discussion

Results from our studies have confirmed Hypothesis 1, a human's cognitive ability is augmented as a result of information supplied by the cog involved in the human/cog ensemble. In these studies, we used just the information that would have been supplied by a cognitive system as hints or guidelines. Study #1 (the innovation challenge) showed students' ability to produce a desired type of solution, cognitive accuracy, was enhanced by 74% and the propensity to produce solutions other than the desired type was enhanced by 27%. Study #2 shows information in the form of rules increased cognitive precision by as much as 42% and information in the form of examples increased cognitive precision by as much as 65%. Furthermore,

Study #2 showed a dramatic increase in cognitive accuracy, 100% for information in the form of rules and 200% in the form of examples.

Cognitive augmentation has been indicated elsewhere. In dermatology, Google's Inception v4 (a convolutional neural network) was trained and validated using dermoscopic images and corresponding diagnoses of melanoma [22]. Performance of this cog against 58 human dermatologists was measured using a 100-image testbed. Measured was the sensitivity (the proportion of people with the disease with a positive result) and the specificity (the proportion of people without the disease with a negative result). Results are shown in Table 3.

**Table 3.** Potential increase in cognitive accuracy and cognitive precision using a neural network for dermatological diagnoses

| Measure | Human | Cog | Improvement |
|---|---|---|---|
| Sensitivity | 86.6% | 95.0% | $\Delta C_A = +9.7\%$ |
| Specificity | 71.3% | 82.5% | $\Delta C_P = +15.7\%$ |

In this case, sensitivity measures the success rate of getting the correct, or desired result (a positive result for someone with the disease) so therefore is a measure of cognitive accuracy. Specificity measures incorrect, or undesired results (a negative result for someone with the disease), so therefore is a measure of cognitive precision. The group of 58 doctors would have performed better if they had used the trained neural network.

In the field of diabetic retinopathy, a study evaluated the diagnostic performance of an autonomous artificial intelligence system, a cog, for the automated detection of diabetic retinopathy (DR) and Diabetic Macular Edema (DME) [23]. The cog exceeded goals in sensitivity and specificity as shown in Table 4.

**Table 4.** Potential increase in cognitive accuracy and cognitive precision using an automated diabetic disorders diagnosis system

| Measure | Goal | Cog | Improvement |
|---|---|---|---|
| Sensitivity | >85.0% | 87.2% | $\Delta C_A = +2.6\%$ |
| Specificity | >82.5% | 90.7% | $\Delta C_P = +9.9\%$ |

Here, sensitivity measures the success rate of getting the correct, or desired result (correctly detecting an abnormality) and therefore is a measure of cognitive accuracy. Specificity measures incorrect, or undesired results (failure to detect abnormality when one is present), so therefore is a measure of cognitive precision. The goals in Table 4 are what are expected from human doctors working in the field. Results show the cog outperforms human doctors. Doctors using the cog would be cognitively augmented.

Cognitive systems will continue to grow in capability. They will continue to achieve higher and higher levels of cognition previously thought to belong only to humans. As a result, the future will see more people using high-level cognitive systems. In fact, Fulbright predicts a cognitive system revolution in which millions of otherwise ordinary people (the masses) achieve *synthetic expertise* by collaborating with cogs via their smartphones, tablets, and standalone devices [14, 24-26]. In these scenarios, people who would normally not be able to achieve expert-level performance will be able to do so because of the interaction with one or more cogs. In the coming decades, Fulbright sees cognitive systems being developed for just about every endeavor affecting peoples' lives much like computers, the Internet, and social media have done over the last few decades. These "synthetic experts" will be cognitively augmented to the expert level.

However, we are at the very beginning of understanding the best ways to achieve such cognitive augmentation.

Cognitive systems acting as our peers and colleagues have been predicted for some time [27, 28]. Author and journalist Walter Isaacson maintains "…the future might belong to people who can best partner and collaborate with computers" [29].